\documentclass[a4paper,UKenglish]{lipics-v2018}

\usepackage{microtype}

\title{100 prisoners and a lightbulb -- looking back}

\author{Vladan Majerech}{Department of Theoretical Computer Science and Mathematical Logic (KTIML), Charles University, Malostransk\'e n\'am\v est\'\i\ 25, Prague 118 00, Czech Republic}{maj@ktiml.mff.cuni.cz}{ https://orcid.org/0000-0003-3006-2002}{}

\authorrunning{V. Majerech}

\Copyright{Vladan Majerech}

\subjclass{Information systems}


\category{}

\relatedversion{}

\supplement{}

\funding{}

\acknowledgements{}

\EventEditors{}
\EventNoEds{1}
\EventLongTitle{}
\EventShortTitle{}
\EventAcronym{}
\EventYear{2020}
\EventDate{}
\EventLocation{}
\EventLogo{}
\SeriesVolume{}
\ArticleNo{1}
\nolinenumbers 
\hideLIPIcs  

\begin{document}
\newcommand{\prevfootnotemark}{\csname @footnotemark\endcsname}
\newcommand{\numfootnote}{\footnote}

\maketitle

\begin{abstract}
100 prisoners and a light bulb is a long standing mathematical puzzle. The problem was studied mostly in 2002 \cite{W2002}, 2003 \cite{DFS2003}, and 2004 \cite{W2004}.
Solutions in published articles had average number of visits above 3850, but 
best solutions on forums had (declared) average number of visits around 3500.
I spent some time in 2007 - 2009 to optimize the communication strategy and I pushed the average number of visits below 3390, seems no new ideas appear after it.
Recently I have met several people familiar with published papers from 2002-2003 but not knowing newer results.
Even after 2009 several papers on the topic were published where the new results were not mentioned \cite{DEW2010}. Whole book was written about the problem \cite{DK2015}.
This is why I am writing this summary.
\end{abstract}

\section{Problem description}
Let us start with one wording of the puzzle:
"100 prisoners are imprisoned in solitary cells. Each cell is windowless and soundproof. There's a central living room with one light bulb; the bulb is initially off. No prisoner can see the light bulb from his or her own cell. Each day, the warden picks a prisoner equally at random, and that prisoner visits the central living room; at the end of the day the prisoner is returned to his cell. While in the living room, the prisoner can toggle the bulb if he or she wishes. Also, the prisoner has the option of asserting the claim that all 100 prisoners have been to the living room. If this assertion is false (that is, some prisoners still haven't been to the living room), all 100 prisoners will be shot for their stupidity. However, if it is indeed true, all prisoners are set free and inducted into MENSA, since the world can always use more smart people. Thus, the assertion should only be made if the prisoner is 100\% certain of its validity.

Before this whole procedure begins, the prisoners are allowed to get together in the courtyard to discuss a plan. What is the optimal plan they can agree on, so that eventually, someone will make a correct assertion?

What is the strategy to beat the warden?"

Note the fixed frequency of visits, so the current visit count is a shared knowledge. 
Note the "certainty approaching 1" is not considered to be enough. To emphasize that I prefer formulation requiring proof based on the communication strategy, where warden logs the history and is able to check the prisoners behaved according to the strategy used in the proof (but the original formulation is definitely more readable).

Forums are flooded by out of box solutions with making marks of any sort in the room, breaking the bulb and so on.
This is why the switch is much more important than the bulb, and the rules allow changes in the room made by warden except changing the switch state. Alternative formulation preventing that would be warden visits a prisoner and gives him 1 bit of information from the previous visit and asks him, 
which bit of information should be given on the next visit.

The problem we solve is $100A2$ version of more general $n$ prisoners and a $k$ state switch problem (with the visit count) denoted $nAk$, problems when the visit frequency/count is unknown are $nBk$ (when the initial switch state is known) resp. $nCk$, (when the initial state is unknown). Both $nBk$ and $nCk$ problems are much easier to solve as the strategies cannot depend on the visit count (and there is a little room for optimization).
We will ignore problems with $k>2$ here. We will concentrate on $100A2$. 

\section{Long known solutions}
In~\cite{W2002} following competitive strategies were mentioned:
Single counter (collector) solution, two level counting scheme, binary counting scheme and dynamic counter selection trick.
Let me describe them using common terminology. 
Only small number of bits is sufficient for description of the states during the process what allowed William Wu to compute expected number of visits explicitly. Unfortunately for more complicated strategies the results could be only approximated. I will mention these strategies as well.

\section{Communication terminology}
To describe a strategy we have to describe two things. One is the meaning of ON/OFF state at $i$-th night ({\bf global signaling} in short). We would use alternatively terms visit count, day and night. After night 0 the game starts on day 1, night 1 follows, than day 2, \dots. A prisoner visiting on day $d$ knows the visit count $d$, he sees the signal from night $d-1$ and leaves in the room signal $d$ according to $d$-th night signaling.

Often the description of global signaling is split to intervals of nights.
The other is individual prisoners strategy, but this strategy at day $i$ is limited by using signals of night $i$. 
Often the individual strategy can be easily deduced from the global signaling.

Let us introduce tokens to simplify description of signaling. 
Imagine each prisoner has a virtual {\bf token}. Goal of our strategy is to let the virtual token in the room to be collected by a token {\bf collector}. When the collector(s) is(are) sure virtual tokens of all prisoners were collected, the game could be terminated (as collecting a prisoners $P$ token ensures the prisoner $P$ was in the central living room). We will omit the word virtual in the rest of the paper, but remember we are strictly following rules and the switch (bulb) state is the only communication method. Prisoners should deduce the tokens from the state according to global signaling. 
I hope negative tokens support the understanding of virtuality.

Let us start with $nB2$ case, where we are restricted such that signaling at all nights is the same. 
One switch state (denote it ON) corresponds to case token being present in the room, 
while the other switch state (denote it OFF) means no token is present there.

To make the scheme work $n-1$ prisoners have to try let their tokens in the room (but they cannot left the room with two tokens as there is no such signal) so their strategy is clear. If they enter in ON state, they should let the state ON and their count of tokens does not change. If they enter in OFF state and they hold a token, they drop it by switching state ON (of course they could not switch it ON if they hold no token).

Before the game starts one prisoner should be chosen to be a collector. His goal is to collect all $n-1$ tokens.
To achieve this goal he can formally start with token count set to $1-n$. When he enters the room in ON state, he can increase the token count and switch OFF. Otherwise he does nothing. When his token count reaches $0$ he knows all required tokens were collected and the game could be terminated.

One collector $nA2$ strategy imitates this $nB2$ strategy. The only difference is the unknown switch state at night $0$, which should be at the start of the first visit changed to OFF, and this is the last time visit count is used. From this time the $nB2$ strategy is followed.

Expected number of visits for the $nB2$ case could be easily calculated. Divide the nights to $n-1$ periods with switch OFF and $n-1$ periods with switch ON. Expected length of period with switch ON is $n$ as only the collector could switch it OFF (and probability of its visit is $1/n$). Expected length of $(n-i)$-th period with switch OFF is $n/i$ as only $i$ prisoners hold  tokens (and probability of visit of one of them is $i/n$). Summing together we got $n(n-1)+nH_{n-1}$ expected number of visits,
where $H_n=1+1/2+1/3+\dots+1/n$ is the harmonic series (and $1>H_n-\ln (n+1)>0$, limit of the difference $\gamma\approx 0.577$ is called Euler–Mascheroni constant).

Problem $nC2$ could not be easily transformed to $nB2$ case as the initial state is not known. This could add one token to the system, so one prisoner may hold his token when $n-1$ tokens are collected --- breaking the proof. One solution lets prisoners start with 2 tokens and terminating when $2n-1$ tokens are collected.
At most one token is hold by a prisoner at the end, what means he visited the central room at least once.
This solution has expected number of visits bigger than $(2n-1)n$ (length of ON phases). 
Length of OFF phases is smaller, but more complicated to express as it depends on how many prisoners hold 2 tokens.

Better $nC2$ strategy lets prisoners start with only 1 token, but possible initial ON state should signal no token. To achieve this, collector on it's first visit does not increase his token count even in the case the state is ON. This requires prisoners not to left their tokens in the room unless the collector visits it the first time. To achieve this, prisoners cannot switch from OFF to ON unless they have seen state ON before. So a (non-collector) prisoner starts in a waiting state and enters $nB2$ strategy state after he sees ON state. Unfortunately the collector strategy has to change once more to prevent a prisoner remaining in wait state all the time. In the case the collector enters in the OFF state, he with nonzero probability decreases his token count and leaves token in the form of ON state in the room. Optimal choices for this nonzero probability 
(parametrized by the game history of the collector) are out of scope of this paper\footnote{I like strategy when the collector computes probability a prisoner with token is in $nB2$ state. He can maintain required statistic using Bayesian tricks. He let the switch OFF when probability a prisoner with token is in $nB2$ state exceeds $1/2$. One should be careful not to underflow (to 0) during the probability computations.}. When there are prisoners with tokens in $nB2$ state, collector enters in ON state much often than in OFF state and prisoners in waiting state have a big chance to switch to $nB2$ state without other influence of the collector. This is why this strategy has only slightly higher expected number of visits than the $nB2$ strategy (the multiplication constant becomes $(1+\varepsilon)$, where $\varepsilon>0$ is much closer to $0$ than to $1$).

Let us concentrate on $nA2$ cases. What are the advantages of possible signaling dependence on the visit count? What are the bottlenecks of the $nB2$ strategy?

The problem is the total time $n(n-1)$ spent waiting in ON state for the collector to switch OFF.
We can either reduce the number of the periods waiting for the switch from ON to OFF state or reduce the length of such a period. Let us start with the former.

Problem with the $nB2$ solution is the initial $1-n$ token count for chosen collector is too far from $0$. Could we do better?

{\bf Dynamic collector selection --- snowball pre-phase} is the way to go.
Let us try to collect as many tokens before naturally choosing the collector.
Let the strategy of all prisoners be the same until the collector is dynamically selected.
Snowball length $\ell$ has to be carefully chosen.
There are $-n$ tokens in the room before the first visit independently on the state.
First entering prisoner discards token and leaves $1-n$ tokens in the room. He sets state ON (just for compatibility reasons). 
State ON at night $g\le\ell$ means there are $g-n$ tokens in the room. 
State OFF at night $2<g\le\ell$ means there are no tokens in the room. 
{\bf Second night exception} defines meaning of state OFF for night $2$ such that there are $1-n$ tokens in the room.
This exception increases probability of collecting one token even in the case for the first two visits the same prisoner was chosen (subtracts $n/(n-1)$ from expected visit count in the single collector case).
At the night $\ell+1$ the normal signaling starts (for single counter state ON means there is 1 token in the room and state OFF means there is no token in the room).

When prisoner enters day $d$ he collects tokens in the room\footnote{Their amount can be expressed using c like ternary conditional (?:) as ((state$\vee d<2$)?($d\le \ell+1$?$(d-1)-n$:$1$):($d=2$?$(1-n)$:$0$))}, 
If he is able to let in the room tokens required by ON signaling and left holding non-negative (0 in this case) tokens, he does so and signals ON.
Otherwise he let signal OFF and corresponding number of tokens in the room. 
During the day 2 this means he would hold 0 tokens and $1-n$ tokens remain in the room, 
other day he would have negative number of tokens and be/become collector and he will left signaling 0 tokens.
When a prisoner with $-1$ token (collector) collects remaining $1$ token, he finishes the process.

This translates that if $d\le\ell$ and he is first time in the room and the state was ON he lefts all tokens in the room and the state remains ON.
If $d=2$ (state must be ON) and he is 2nd time in the room, he switches state OFF (and let tokens in the room intact).
If $\ell\ge d>2$, state is ON and he is 2nd time in the room, he switches OFF, grabs all the ($(d-1)-n$) tokens in the room and becomes collector.
If $d=3$ and state is OFF, he grabs all the $(1-n)$ tokes in the room and becomes collector.
If $\ell\ge d>3$ and state is OFF, the collector was already chosen and prisoners keep their tokens and signal remains OFF.
If $d=\ell+1$ and state is ON, he becomes collector, he grabs $\ell-n$ tokens from the room and switches OFF.
At the moment $d=\ell+1$ and state is OFF, the normal signaling starts. 
A prisoner with positive number of tokens (1) tries to discard the token by switching from OFF to ON, a prisoner with 0 tokens does not change the state, 
and the prisoner with negative number of tokens (collector) tries to grab tokens by switching from ON to OFF and the collector 
declares process termination when he reaches count 0 of tokens.

Each token collected during the snowball decreases the expected number of visits by $n+n/i$ (with corresponding $i$). To optimize the expected visit count $\ell$ has to be chosen by the following logic: If probability the collector is not chosen yet ($n^{-\ell}n!/(n-\ell)!$) multiplied by decrease of expected number of visits if we collect one more token ($n+n/(n-\ell)$) is less than 1, snowball should not continue. According to Stirling approximation the probability collector is not chosen yet is $(n^{-\ell}n^n/(n-\ell)^{n-\ell})(e^{n-\ell}/e^n)(\sqrt {2\pi n}/\sqrt {2\pi (n-\ell)})\cdot c$ where $e^{1/(12n+1)}/e^{1/(12(n-\ell))}<c<e^{1/12n}/e^{1/(12(n-\ell)+1)}$.
We can adjust it to form $(n/(n-\ell))^{n-\ell}e^{-\ell}\sqrt{n/(n-\ell)}c=(1+\ell/(n-\ell))^{n-\ell}e^{-\ell}\sqrt{n/(n-\ell)}c$. 
Let us adjust bounds for $c$ as well $1/(12n+1)-1/(12(n-\ell))<\ln c<1/12n-1/(12(n-\ell)+1)$ so $-{l+1\over 144n^2+(1/12-\ell)n-\ell/12}<\ln c<{1/12-\ell\over 12n^2-11n}$.
So we have to compare $(n+n/(n-\ell))\cdot \sqrt{n/(n-\ell)}(n/(n-\ell))^{n-\ell}e^{-\ell}c$ with 1 or equally $\ln (n+n/(n-\ell))+(n-\ell+1/2)\ln (n/(n-\ell))+\ln c$ with $\ell$.
This gives for example $\ell(100)=29$, $\ell(10000)=426$ and $\ell(1000000)=5252$.

\bigbreak 

Let us continue with the latter. 
Let us reduce the length of period in the ON state we have to increase the probability the chosen prisoner would turn OFF.
We can achieve this only if more prisoners would be able to switch OFF.
First solution uses {\bf multiple collectors} scheme. 
In this scheme several collectors are chosen and $i$-th of them subtracts $g_i$ from his token count to reflect his {\bf goal} to collect $g_i$ tokens. It should hold that $n=g_1+g_2+\dots$ so the total goal of all collectors is to collect all tokens. While there are $t$ collectors with negative token count, the expected length of period with ON state is just $n/t$. So the expected length of such phases is reduced by almost factor of $t$, while the length of remaining phases remain almost same. This unfortunately is not for free. 
We have to check all collectors finished their job, but token signaling is not good for that. Instead we introduce {\bf talent} signaling. One of collectors is chosen as head collector and other collectors signal their finished jobs by sending talents when possible. 
We have to reserve some nights for tokens signaling and other nights for talent signaling ({\bf token phase/talent phase}). ON means signal of the corresponding type is present in the room. Unfortunately optimizing the phase lengths is not easy. 
Continuing token phase when all tokens were collected is waste of time, stopping the phase when they were not collected means the talent phase cannot not be successful and another token phase should be started later. 
First token phase should last so long the probability all the tokens were collected approaches 1.
First talents phase should last so long the probability all the talents were collected (even when all talents are ready for collection) approaches 1.
Following recovery phases should last at least $2n$, what is expected number of visits when just one token resp. talent was not collected.
The phase switch means the signal from the last night of the phase should be collected by the prisoner in the room even when it is not his intention. (For example non-collector could obtain talent at the end of talent phase. He will try to pass it during the next talent phase. Similarly unlucky prisoner could end up with two tokens at the end of first token phase or with astronomically low probability an unlucky prisoner could end up with $n-1$ tokens at the end of $n-1$-st token phase).

Pre-phases are cheap as there is no {\bf phase switch risk}, but switching to recovery phases is expensive. 
Let us lower bound the expected number of visits assuming there are no visits lost to the phase switching. In token phase there are $(n-t)$ periods with state OFF and $(n-t)$ periods with state ON. The expected number of visits in the former is $nH_{n-t}$, the expected number of visits in the latter is at least $(n-2t)n/t+nH_t$. In talent phase there are $t-1$ periods with state OFF and $t-1$ periods with state ON. Expected number of visits of the former is $nH_{t-1}$, expected number of visits of the latter is $nt$. Lower bound of expected number of visits is therefore $n(H_{n-t}+H_t+H_{t-1}+n/t-2+t)$. Asymptotically optimal choice is $t\approx\sqrt{n}$ giving $\theta(n^{3/2})$.

{\bf Multiple snowballs} is the way to choose multiple collectors. Choosing snowballs parameters is difficult. Good choice is to set the first goal be biggest as the first snowball has highest expected number of collected tokens. It is crucial to choose different prisoners for each goal with high probability.
The parameters could be described by a tuple $((\ell_1,s_1,g_1),(\ell_2,s_2,g_2),\dots,(\ell_t,s_t,g_t))$ with $g_1+g_2+\cdots+g_t=n$. 

Let us define the signaling first (we will use notation $p_k=(\ell_1+s_1)+(\ell_2+s_2)+\cdots+(\ell_k+s_k)$):
\begin{center}
\begin{tabular}{| c | r | r | r |}
\hline
state & night $g$ & tokens & talents \\
\hline
any & $g=0$ & $-n$ & $0$ \\
ON  & $1\le g\le\ell_1$ & $g-n$ & $0$ \\
OFF & $g=2$ & $1-n$ & $0$ \\
OFF & $2<g\le p_1$ & $g_1-n$ & $t-1$ \\
ON  & $\ell_1<g\le p_1$& $\ell_1-n$ & $0$ \\
ON  & $p_k<g\le p_k+\ell_{k+1}$ & $(g-p_k)+(g_1+\cdots+g_k)-n$ & $t-k$ \\
OFF & $g=p_k+1$& $(g_1+\cdots g_k)-n$ & $t-k$ \\
OFF & $p_k+1<g\le p_{k+1}$& $(g_1+\cdots g_{k+1})-n$ & $t-(k+1)$ \\
ON  & $p_k+\ell_{k+1}<g\le p_{k+1}$& $\ell_{k+1}+(g_1+\cdots+g_k)-n$ & $t-k$ \\
OFF & $p_t<g$ & $0$ & $0$ \\
ON  & $p_t<g$ token phase& $1$ & $0$ \\
ON  & $p_t<g$ talent phase& $0$ & $1$ \\
\hline
\end{tabular}
\end{center}

While the snowball subphase is not interrupted, each day the number of tokens in the room increases by 1.
Signal ON means current snowball subphase was not interrupted yet. Signal OFF means the opposite. 
When the interruption occurs in the first possible night, the signal OFF means the next entering prisoner should become collector for the subphase\footnote{A collector would restart it by dropping two tokens rather to becoming a double collector.}.
When the interruption occurs later, OFF signal cannot carry the information, so the collector should be chosen immediately.
To become a collector for the phase $k+1$ the prisoner lefts $g_1+g_2+\cdots+g_{k+1}-n$ tokens in the room. It effectively means he lefts $g_{k+1}$ tokens in the room 
and takes all tokens collected during the subphase $k+1$. Becoming collector of the first snowball leaves $t-1$ talents in the room, 
while becoming a collector of another subphase takes 1 talent.
A bit complicated is the behaviour on the boundaries of subphases.
At day $d$, where $p_k+\ell_{k+1}<d\le p_{k+1}$ if the snowball was not interrupted yet and the prisoner has 0 talents, he should interrupt the subphase and become its collector.
Note that a collector chosen in previous snowballs does not left a token here as his snowball could have been interrupted early, while current is not.
At day $d=p_{k+1}+1$ if the signal is ON the prisoner should interrupt the subphase $k+1$ and become its collector. Moreover, if another snowball continues, 
he does not want to become its collector as well, so he lefts a token in the room.

The behaviour of prisoners without talents is straightforward. The behaviour of prisoners with talents is more complicated.
They could choose to restart already interrupted subphase. 
This of course means they should left talents and tokens in the room compatible with the global signaling.
They certainly do it to prevent to have more than 1 talent. 
But they could do it even in the case they have just 1 talent if the expected number of tokens collected during the rest of the snowball subphase is bigger than the number of tokens he should keep. He cannot restart the snowball subphase if it would mean ending with negative number of tokens and 0 talents. Note that restarts could lead to situation talent collector is not token collector so he starts with $-t$ talents in that case (this would increase the expected visit count so it has to be included in restart choice effectivity calculation).
 
The token total and talent total are all the time equal 0. Note that having $s_1>0$ is pointless\footnote{$(\ell-1,1,g)$ differs from $(\ell,0,g)$ only when first $\ell$ visitors are diferent and one of first $\ell-1$ enters day $\ell+1$. In latter signaling the collector is able to discard one token not to interrupt the next snowball subphase.}, experimental results suggests having $s_k>1$ is not optimal\footnote{$(\ell_k-1,1,g_k)$ differs from $(\ell_k,0,g_k)$ only when state was ON at night $p_{k-1}+\ell_k-1$ and the prisoner entering has 1 token and 0 talents. When next day a collector comes he becomes double collector in the latter case. If a prisoner with 0 tokens enters (much often), the next snowball prephase is interrupted in the former case. It is not easy to guess what variant gives smaller expected visit count.}.
Our goal is to get at most $t$, but as close to $t$ as possible collectors with roughly equal negative token count (with the sum of these counts as close to 0 as possible) at the end of the pre-phases.

During talent phases a collector with non-negative number of tokens tries to leave the talent in the room. Collector with more than 1 talent tries to leave talent in the room as well.
When a talent collector reaches 0 talents and 0 tokens, the game is terminated. All other collectors left their last talent in the room declaring they collected their goal of tokens. As the sum of goals is $n$ this means all non-collector prisoners left their tokens in the room so all prisoners visited the room.

Note the snowballs are able to decrease multiplicative constants, but they cannot affect the asymptotic\footnote{When nontrivial portion of prisoners is without tokens, the number of collected tokens by snowballs become negligible. This is why nontrivial portion of prisoners should remain with tokens at the end of snowballs.}. The first fail tricks subtract from expected number of visits about one day rather than two. Similarly the influence of $s_k=1$ is negligible and guessing optimal setting by a random experiment is inconclusive.

\bigbreak 

Another try to shorten expected number of visits during the ON state period is allowing more active prisoners to switch OFF. 
There is a {\bf binary schema} with signaling $2^k$ tokens (or talents) in appropriate stage. Each prisoner with number of tokens/talents having 1 in $k$-th bit is active during the stage. Passive (nonactive) prisoners do not change the switch state during the stage, while active prisoners always do. 
This means when an active prisoner enters and switches state from OFF to ON, he has to leave $2^k$ tokens/talents in the room. Changing state from ON to OFF means collecting $2^k$ tokens/talents.
In either case the prisoner becomes passive for this stage.
The total expected number of visits to make all $a$ active prisoners passive is $nH_a$.
Unfortunately this strategy requires $\log_2 a$ successful stages to collect $a$ tokens/talents by one prisoner.
The phase switching problems are more severe the more stages could end in unfinished state.

There is a small problem when the total number of tokens in the prison is not in the form $2^k$.
The problem can be solved by adding number of tokens required to round to nearest power of 2 to the room before the first visit. When the first phase with $2^k$ signaling starts, the state is changed accordingly to allow collecting possible token amount during the stage.

Lower bound on expected number of visits (ignoring the phase switching) is $n(H_n+H_{n/2}+H_{n/4}+H_{n/8}+\cdots +H_1)\in\theta(n\log^2 n)$ for token only collection.

Experiments show that for $n=100$, where $\sqrt{100}<\log^2{100}$, the higher state switching cost and lowering constants by multiple snowballs the two level collectors scheme behaves better than the binary token counting scheme.

As snowballs are efficient in reducing the multiplicative constants, binary scheme for collecting talents are promising way to go.
Unfortunately the phase switching/error recovery problems are more severe and single talent collector strategy still works better for $n=100$.
Note the snowballs does make prisoners on level $2^0$ inactive less efficiently than binary scheme, so combining it with token binary scheme could help only 
in later stages, but even there the eventual help will be negligible.

There is also collector binary accelerated strategy in which token/talent collector grabs tokens/talents in corresponding $2^i$ chunks, and active prisoners join the chunks as in binary schema. The strategy does not necessarily use all possible binary levels and in repair phases the maximal used level decreases (what could result in splitting some chunks). 
Reducing the number of different signaling decreases the phase switch cost in later stages, unfortunately the phase switch cost in early phases is high. Note that token version could easily be combined with a snowball.
I have not experimented with it.
Expected number of visits is asymptotically $\theta(n\log^2 n)$ for token variant as for binary schema.
The help of active prisoners decreases when there is small number of them. This is probably why in experiments binary talent observers perform better then collector binary accelerated talents.

Let us approximate the snowball length for binary accelerated strategy. Gain by prolonging running snowball is between $n/(n-\ell)$ and $n+n/(n-\ell)$ let us expect it is about $2n/(n-\ell)$ so we have to compare $2n/(n-\ell)\cdot \sqrt{n/(n-\ell)}(n/(n-\ell))^{n-\ell}e^{-\ell}c$ with 1 or equally $\ln2+(n-\ell+3/2)\ln (n/(n-\ell))+\ln c$ with $\ell$.
This gives informed guesses $\ell(100)=13$, $\ell(10000)=119$ and $\ell(1000000)=1179$. Optimum is somewhere between these bounds and bounds for single collector case.

\section{Observers}
From the so far described communication schemes one could get a feeling only few prisoners have fun during the game. The prisoners without tokens become passive visitors of the central room. This is actually true only for one collector schema (possibly with the snowball). 
When there are more collectors, each prisoner could make statistics of ON OFF observations. In the unlikely case he sees $n$ changes from ON to OFF states during a token collection phase, he can declare termination. This is very unlikely especially when snowballs are used, so the prisoners would not be motivated enough to do such activity.
The binary case is different. With the total number of $2^k$ tokens, observation at the phase with $2^{k-2}$ signaling is very promising. 
Seeing ON OFF ON OFF during the phase have big enough probability to be worth of making statistics. 
Such observation could terminate the game even when the stage collecting $2^{k-1}$ tokens did not started.
With much smaller probability the game could be terminated already in the phase with $2^{k-3}$ signaling.
Actually even the active prisoners could make their statistics and the original win condition (joining two bunches of $2^{k-1}$ tokens) could be interpreted as a special case of observer stop (seeing ON OFF during the phase). In this interpretation the original strategy is just a game not declaring termination, but allowing observers to declare it.

Does observers game influence the expected number of visits by a measurable amount?
I do not think so, unless we make some other changes.

{\bf Binary observers statistics} include the state and night count he last left the room and tuple $(o_0,o_1,\dots,o_k)$, 
where $o_0$ is the number of tokens/talents he knows were signaled and $o_i$ for $i>1$ is the number of tokens/talents he knows were ever ready for level $i$ signaling.
The statistic is defined such that $o_0\le o_1\le \cdots\le o_k$.
Whenever he increases $o_i$, he makes sure $o_{i-1}$ is not smaller (and increases it if required). 
Observation of ON during $2^i$ signaling means $o_i\ge o_{i+1}+2^i$ (so increases when required). 
Observation of OFF during $2^i$ signaling has no meaning when last observation was not in the same $2^i$ phase. 
But if the observation was in the same phase and it was ON, he adds $2^{i+1}$ to $o_{i+1}$ as he observed join of two $2^i$ chunks of tokens/talents.
When $o_0$ reaches the number of tokens/talents in the system, the game can be terminated. 
Note that a prisoner in the room at day $d$ observes the state of both $d-1$ and $d$ night ($d$ and state from the night $d$ is remembered).
The signaling caused by extra tokens from the start should be taken into account as well.
Consider 16+8+4 extra tokens for the $100A2$ case. When a prisoner enters after last night of first $2^1$ signaling, if state is ON, he collects 2 tokens, he set switch to state ON and 
the 4 tokens are signalled by it (they are no more extra-tokens). Each prisoner sets the last observation is from the start of first $2^2$ signaling and the state is ON and adds $4$ to $o_2$ (it has independent source than other known $2^2$ tokens). If the prisoner in the room is $2^2$ active, he can switch OFF and take the tokens, so the signal at the first signaling night could be OFF (what would create $o_3=8$ observation of both the active prisoner and the prisoner entering next day.
Similarly the 8 and 16 tokens are released. 

Typically when a phase fails, there is very few amount of work left in the phase. Following phases cannot finish either. 
In binary scheme typically only one super-token per level is not created. What means signalization of $2^{k-1}$ tokens ended with long ON period, so almost all observers know the 
$2^{k-1}$ goal is fulfilled ($o^p_{k-1}=2^{k-1}=2^k-2^{k-1}\le o^p_{k-2}\le\cdots\le o^p_0$).

If the recovery stages with signaling $2^{k-2}$ follows, it typically ends with long ON period as well, so almost all observers know both $2^{k-1}$ and $2^{k-2}$ goals were independently fulfilled ($o^p_{k-2}=2^k-2^{k-1}+2^{k-2}=2^k-2^{k-2}\le o^p_{k-3}\le\cdots\le o^p_0$). 
So improved strategy changes first cycle of recovery stages to have signaling $2^{k-2}$, $2^{k-3}$, \dots, $1$. 
Typically recovery stage with signaling equal to the signaling from the fist stage which was not successfully finished is the last started phase as observers require info only about the missing amount of tokens.

The more efficient recovery, the more risk could be allowed, so the optimal stage lengths are shorter.

The token binary scheme with observers and optimized order of stages become competitive with two level counting even for small $n$. The multi-snowball multi-collectors with talent binary observers scheme got better results for small $n$ than both two level counting and binary token schema.
It seems the phase switch problem is dominated by the recovery strategy caused by binary observers.

The following sections would discuss method used in parameter tuning and in the experimental evaluation of the strategies.

Surprisingly from the simulations another strategy was born. 
The optimal number of talents seems to be $3\cdot 2^k$ rather to $2^k$. 
This is incompatible with binary scheme without observers, but observers method likes it (I call it ternary observers).
For $100A2$ problem seems this schema with 12 talents and well chosen parameters is best known strategy so far.

According to asymptotic analysis, token binary observers or one collector with snowball combined with binary accelerating schema should be optimal strategy for big $n$.
Interesting question, I have not simulated is which of the strategies prevail for huge $n$ and for which $n$ 
they start beating $3\cdot 2^k$ talent binary observers with multiple snowballs.

\section{Parameter optimization method}
Each of mentioned communication protocols could be characterized by lengths of its phases (with specified signaling) and (possible snowball characterizing tuple).
For example the $100B2$ protocol with snowball of length 29 could be described by $((29,0,100),\infty,K)$.
When we are restricted on a protocol of given type, we know the signaling and it need not be included in the description. 
As we are describing infinite protocols, there should be a pattern determining the lengths of all phases. 
We typically let the recovery phases starting from 2nd to have equal parameters as they have negligible influence on the expected visit count.
I made a decision to include pre-phases in the first phase length. So changes in snowball parametrization has no influence on the visit count, when the first phase ends.
When binary phases and collector phases are used in the algorithm, I have chosen to maintain the lengths of collector phases separately from lengths of binary phases.
The reason is expected visit count for one fail recovery is $2n$ for collector phases while $3n/2$ for binary phases so the repeated last time should be parametrized differently.

I have not found method how to find optimal parameter settings so I have made an informed guess for each method and I started simulations.
I decided to evolve the parameters by following strategy. For each parametrization, I have remembered number of simulations and total visit count of these simulations.
With small probability I have modified the parametrization and I have stored the results based on the parametrization. To decide which parametrization (gen) to start with, I have generated random number $r$ in interval $(0,1)$, if there were $s$ simulations so far, I have chosen in list of parametrizations (gens) ordered by increasing average number of visits per gen, the gen when total count of simulations exceeded $r^3s$.
This favors more successful gens. Gen changes were not fully random, but reflected expected patterns 
(there is no reason to have $g_{i+1}>g_i$ or $s_i>s_{i+1}$, first collectors phase should be longer than recovery collector phases, \dots).

I wanted to find the limits of individual methods so I have simulated each main strategy separately. So I have obtained averages for gens with two level collectors with multi snowball for different $t$, averages for gens with $2^k$ or $3\cdot 2^k$ collectors with multi snowball and with binary/ternary observers for different $k$.

I have reported average of all simulations of all gens (population average). It should be upper bound of the average of the best gen. As gens with worse average were selected less often, they did not influence the result too much. I was remembering per gen total number of simulations and total number of visits I made snapshots of these totals ocassionally and this allowed me to calculate average from the last snapshot time (last generation average).
I have used Mersene twister as a pseudorandom numbers generator.

Unfortunately I cannot find the source codes, so I will present only the old results.
Will I or anyone else recode and repeat the experiments?

\begin{figure}
\begin{tabular}{| c | r | l |}
\hline
$n$ & average number of visits $g$ & algorithm \\
\hline
$2$ & $3$ & binary (discard first token)\\
$3$ & $5.5$ & binary, discard first token\\
$4$ & $12.3$ & ternary, discard first token\\
$5$ & $\approx 23.26$ & $((3,0,5),\infty)$ single collector with 3 nights snowball\\
$6$ & $\approx 33.76$ & $((4,0,6),\infty)$\\
$7$ & $\approx 45.90$ & $((4,0,7),\infty)$\\
$8$ & $\approx 60.01$ & $((5,0,8),\infty)$\\
$9$ & $\approx 75.72$ & $((5,0,9),\infty)$\\
$10$ & $\approx 93.39$ & $((5,0,10),\infty)$\\
$11$ & $\approx 112.76$ & $((6,0,11),\infty)$\\
$12$ & $\approx 133.97$ & $((6,0,12),\infty)$\\
$13$ & $\approx 157.05$ & $((7,0,13),\infty)$\\
$14$ & $\approx 181.82$ & $((7,0,14),\infty)$\\
$15$ & $\approx 208.52$ & $((7,0,15),\infty)$\\
$16$ & $\approx 233.0$ & $((6,0,9)(3,1,7),(166,193)()(82)()$\\
$17$ & $\approx 257.6$ & $((5,0,9)(3,1,8),(182,173)()(96)()$\\
$18$ & $\approx 292.7$ & $((5,0,10)(4,1,8),(201,214)()(126)()$\\
$19$ & $\approx 307.9$ & $((5,0,10)(5,1,9),(229,196)()(100)()$\\
$20$ & $\approx 341.7$ & $((5,0,10)(4,1,10),(251,242)()(125)()$\\
\hline
\end{tabular}

\kern2mm
The translation of parameters of strategy for $n=16$ is two collectors with given multi snowball
with binary observers 166-(6+4) nights of token signaling, 
than repeated (82 nights of talent binary signaling, and 193 nights of token signaling).

\kern2mm
\hbox{Figure 1: Results for small $n$.}

\kern2mm
\end{figure}

\section{Problem with variable number of prisoners + simulation results}
I have simulated $nA2$ problems for $n\le 100$. For $n<16$ the optimal solution is probably known, for $16\le n\le 20$ we have 
good candidates, but the optimal parametrization could differ (see figure 1).

For bigger $n$ I have not presented parametrizations, but figure 2 shows the results of genetic algorithm. 
I have run the genetic algorithm for each $n$ and algorithm type separately, I have used scaled parametrization from nearby $n$ as an informed guess to init the simulation. 
I have presented population average and last generation average. As the results are close to $b(n)={\pi\over 2} n\ln^2 n$, I have subtracted $b(n)$ from the averages.
Running simulation longer lowered both the averages and last generation average got closer to the whole population average. This can be seen on peak for $n=89$ for 8 talent observers.

\begin{figure}
\vbox to 100mm{\vss
\hbox{\pdfximage width 14cm {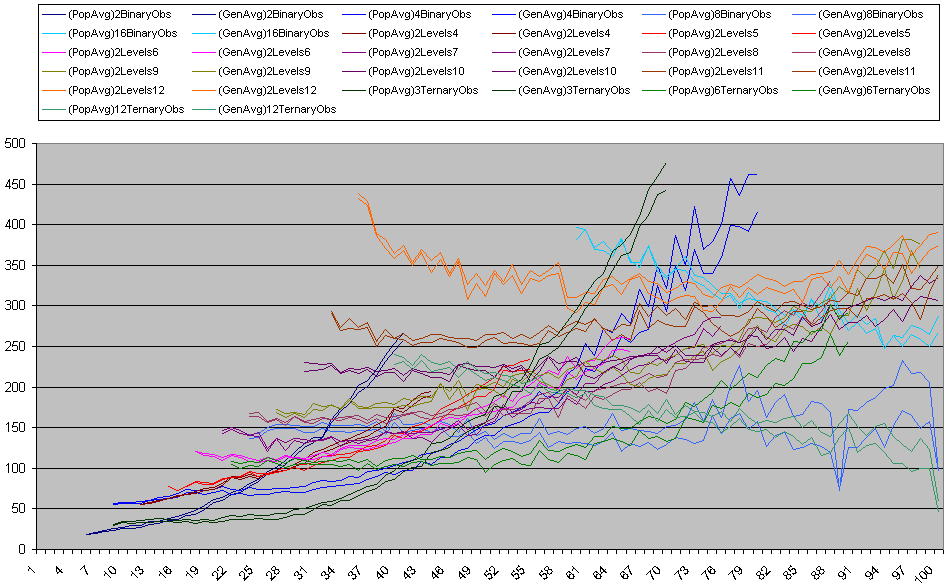}
\rlap{\smash{\pdfsave\pdfsetmatrix{1 0 0 1}
\pdfrefximage\pdflastximage}}\pdfrestore
\hss}
\hbox{Figure 2: Results of genetic algorithms (${\pi\over 2} n\ln^2 n$ subtracted)}
\kern4mm
}
\end{figure}

The best results for 100A2 case are obtained by talent ternary observers with 12 collectors selected by multi snowball. I got under 3390 average of all simulations, most spread gen:
$((3,0,9),(3,1,9)^3,(3,1,8)^8,(1928,657,665),(440,442,439),(506,512),(378))$ had average around 3355. It started with 12 snowball prephases of total length 47 followed by 1928-47 nights of token signaling, followed by 440 nights of 1 talent signaling, 442 nights of 2 talent signaling, 506 nights of 4 talent signaling, 378 nights of 2 talent signaling, 378 nights of 1 talent signaling, 657 nights of token signaling, 439 nights of 1 talent signaling, 439 nights of 2 talent signaling, 512 nights of 4 talent signaling, and forever repeated (665 nights of token signaling, 439 nights of 1 talent signaling, 439 nights of 2 talent signaling and 512 nights of 4 talent signaling). I do not expect 516 nights of 2nd 4 talent signaling is good choice, but 2nd 4 talent signaling does not occur too often to have a noticeable influence on the average.

\bibliography{100palb}

\begin{thebibliography}{1}

\bibitem{DFS2003}
Paul-Ollivier Dehaye, Daniel Ford, and Henry Segerman.
\newblock One hundred prisoners and a lightbulb.
\newblock {\em Methematical Intelligencer}, 25(4):53--61, 2003.
\newblock URL: \url{https://doi.org/10.1007/BF02984862}.

\bibitem{DK2015}
Hans~van Ditmarsch and Barteld Kooi.
\newblock {\em One Hundred Prisoners and a Light Bulb}.
\newblock Copernicus, USA, 1st edition, 2015.

\bibitem{W2004}
A~K Peters.
\newblock Mathematical puzzles: A connoisseur's collection.
\newblock pages 109--111, 2004.

\bibitem{DEW2010}
Hans van Ditmarsch, Jan van Eijck, and William Wu.
\newblock One hundred prisoners and a lightbulb - logic and computation.
\newblock In Fangzhen Lin, Ulrike Sattler, and Miroslaw Truszczynski, editors,
  {\em Principles of Knowledge Representation and Reasoning: Proceedings of the
  Twelfth International Conference, {KR} 2010, Toronto, Ontario, Canada, May
  9-13, 2010}. {AAAI} Press, 2010.
\newblock URL: \url{http://aaai.org/ocs/index.php/KR/KR2010/paper/view/1234}.

\bibitem{W2002}
William Wu.
\newblock 100 prisoners and a light bulb.
\newblock 2002.
\newblock URL:
  \url{http://www.ocf.berkeley.edu/~wwu/papers/100prisonersLightBulb.pdf}.

\end{thebibliography}
\end{document}